\documentclass[twoside]{article}
\usepackage{Proc_NTSE_14x}
\usepackage[english]{babel}

\newcommand{\bbR}{{\mathbb R}}

\newcommand{\cO}{{\mathcal O}}
\newcommand{\sE}{{\sf E}}
\newcommand{\fH}{\mathfrak{H}}
\newcommand{\fP}{\mathfrak{P}}
\newcommand{\fQ}{\mathfrak{Q}}
\newcommand{\dist}{\mathop{\rm dist}}

\newcommand{\Ran}{\mathop{\mathrm{Ran}}}
\DeclareMathOperator{\spec}{spec}

\DeclareSymbolFont{SY}{U}{psy}{m}{n}
\DeclareMathSymbol{\emptyset}{\mathord}{SY}{'306}

\begin{document}
\thispagestyle{plain}
\publref{Albeverio-Motovilov}

\begin{center}
{\Large \bf \strut
Bounds on variation of the spectrum and\\
spectral subspaces of a few-body Hamiltonian
\strut}\\
\vspace{10mm} {\large \bf

Sergio Albeverio$^{a}$ and Alexander K. Motovilov$^{b}$}
\end{center}

\noindent{{\small $^a$\it Institut f\"ur Angewandte Mathematik and
HCM,
Universit\"at Bonn, Endenicher Allee 60, \\
\phantom{\small $^a$}{\small\it 53115 Bonn, Germany}\\
\small $^b$\it Bogoliubov Laboratory of Theoretical Physics, JINR,
Joliot-Curie 6, 141980 Dubna, Russia}

\markboth{
S. Albeverio and A. K. Motovilov}
{
Bounds on variation of the spectrum and spectral subspaces}

\begin{abstract}
We overview the recent results on the shift of the spectrum and norm
bounds for variation of spectral subspaces of a Hermitian operator
under an additive Hermitian perturbation. Along with the known
results, we present a new subspace variation bound for the generic
off-diagonal subspace perturbation problem. We also demonstrate how
some of the abstract results may work for few-body Hamiltonians.
\\[\baselineskip]
\end{abstract}

\section{Introduction}

In this short survey article we consider the problem of variation of
the spectral subspace of a Hermitian operator under an additive
bounded Hermitian perturbation. It is assumed that the spectral
subspace is associated with an isolated spectral subset and one is
only concerned with the geometric approach originating in the papers
by Davis \cite{D63,D65} and Davis and Kahan \cite{DK70}. In this
approach, a bound on the variation of a spectral subspace usually
involves just two quantities:  the distance between the relevant
spectral subsets and a norm of the perturbation operator. We discuss
only the a priori bounds, that is, the estimates that involve the
distance between complementary disjoint spectral subsets of the
unperturbed operator (and none of the perturbed spectral sets is
involved). In the case where the perturbation is off-diagonal, we
also recall the bounds on the shift of the spectrum.

The paper is organized as follows. In Section \ref{Sec2} we collect
the results that hold for Hermitian operators of any origin. Along
with the older results we present a new bound in the general
off-diagonal subspace perturbation problem that was not published
before. In Section \ref{Sec-FB} we reproduce several examples that
illustrate the meaning of the abstract results in the context of
few-body bound-state problems.

In this paper we only use the usual operator norm. For
convenience of the reader, we recall that if $V$ is a bounded linear
operator on a Hilbert space $\fH$ then its norm may be computed by
using the formula
$$
\|V\|=\sup\limits_{f\in\fH,\,\|f\|=1} \bigl\|V|f\rangle\bigr\|
$$
where sup denotes the least upper bound. Thus, one has
$\bigl\|V|f\rangle\bigr\|\leq \|V\|\,\|f\|$ for any
{$|f\rangle\in\fH$}. If {$V$} is a Hermitian operator with
$\min\bigl(\spec({V})\bigr)=m_V$ and
$\max\bigl(\spec({V})\bigr)=M_V$ where $\spec(V)$ denotes the
spectrum of $V$, then $\|V\|=\max\{|m_V|,|M_V|\}$. In particular, if
$V$ is separable of rank one, i.e. if
$V=\lambda|\phi\rangle\langle\phi|$ with $|\phi\rangle\in\fH$,
$\|\phi\|=1$, and $\lambda\in\bbR$, then $\|V\|=|\lambda|$. Another
simple but important example is related to the case where
$\fH=L_2(\bbR^n)$, $n\in\mathbb{N}$, and $V$ is a bounded local
potential, that is, $\langle x|V|f\rangle=V(x)f(x)$ for any
$|f\rangle\in L_2(\bbR^n)$, with $V(\cdot)$ a bounded function from
{$\bbR^n$} to $\mathbb{C}$. In this case
{$\|V\|=\sup\limits_{x\in\bbR^n}|V(x)|$}.

\section{Abstract results}
\label{Sec2}

Let $A$ be a Hermitian (or, equivalently, self-adjoint) operator on
a separable Hilbert space $\fH$. It is well known that if $V$ is a
bounded Hermitian perturbation of $A$ then the spectrum of the
perturbed operator $H=A+V$ lies in the closed $\|V\|$-neighborhood
$\cO_{\|V\|}\bigl(\spec(A)\bigr)$ of the spectrum of $A$ (see, e.g.,
\cite{Kato}). Hence, if a subset $\sigma$ of the spectrum of $A$ is
isolated from the remainder $\Sigma=\spec(A)\setminus\sigma$, then
the spectrum of $H$ also consists of two disjoint components,
\begin{equation}
\label{omOMd} \omega=\spec(H)\cap\cO_{\|V\|}(\sigma)\text{\, and\, }
\Omega=\spec(H)\cap\cO_{\|V\|}(\Sigma),
\end{equation}
provided that
\begin{equation}
\label{Vd12In}
 \|V\|<\text{\small$\frac{1}{2}$}\,d,
\end{equation}
where
\begin{equation}
\label{separIn} d:=\dist(\sigma,\Sigma)>0.
\end{equation}
Under condition \eqref{Vd12In}, the separated spectral components
$\omega$ and $\Omega$ of the perturbed operator $H$ may be viewed as
the result of the perturbation of the respective disjoint spectral
subsets $\sigma$ and $\Sigma$ of the initial operator $A$.

Let $P$ and $Q$ be the spectral projections of the operators $A$ and
$H$ associated with the respective spectral sets $\sigma$ and
$\omega$, that is, $P:=\sE_A(\sigma)$ and $Q:=\sE_H(\omega)$. The
relative position of the perturbed spectral subspace $\fQ:=\Ran(Q)$
with respect to the unperturbed one, $\fP:=\Ran(P)$, may be studied
in terms of the difference $P-Q$ and, in fact, the case where
$\|P-Q\|<1$ is of particular interest. In this case the spectral
projections $P$ and $Q$ are unitarily equivalent and the
transformation from the subspace $\fP$ to the subspace $\fQ$ may be
viewed as the direct rotation (see, e.g. \cite[Sections 3 and
4]{DK70}). Furthermore, one can use the quantity
$$
\theta(\fP,\fQ)=\arcsin(\|P-Q\|),
$$
as a measure of this rotation. This quantity is called the maximal
angle between the subspaces $\fP$ and $\fQ$. For a short but concise
discussion of the concept of maximal angle we refer to \cite[Section
2]{AM-CAOT-2013}; see also \cite{DK70,KMM5,MotSel,Se2013-1}. If
\begin{equation}
\label{tlp2} \theta(\fP,\fQ)<\text{\small$\frac{\pi}{2}$}
\end{equation}
and, thus, $\|P-Q\|<1$, the subspaces $\fP$ and $\fQ$ are said to be
in the acute-angle case.

Among the problems being solved in the subspace perturbation theory,
the first and rather basic problem is to find an answer to the
question on whether the requirement \eqref{Vd12In} is sufficient for
the unperturbed and perturbed spectral subspaces $\fP$ and $\fQ$ to
be in the acute-angle case, or, in order to ensure \eqref{tlp2}, one
has to impose a stronger condition $\|V\|<c\,d$ with some
$c<\frac{1}{2}$. More precisely, the question is as follows.
\begin{enumerate}
\item[(i)] What is the largest possible constant
$c_*$ in the inequality
\begin{equation}
\label{Vcsd} \|V\|<c_*\,d
\end{equation}
securing the subspace variation bound \eqref{tlp2}?
\end{enumerate}
\noindent Another, practically important question is about the
largest possible size of the subspace variation:
\begin{enumerate}
\item[(ii)]  What function $M: [0,c_*)\mapsto \bigl[0,\frac{\pi}{2}\bigr)$
is best possible in the bound
\begin{equation}
\label{Mquest} \theta(\fP,\fQ)\leq
M\left(\text{\small$\frac{\|V\|}{d}$}\right)\quad\text{for}\quad
\|V\|<c_*\,d?
\end{equation}
\end{enumerate}
Both the constant $c_*$ and the function $M$ are required to be
universal in the sense that they should work simultaneously for all
Hermitian operators $A$ and $V$ for which the conditions
\eqref{Vd12In} and \eqref{separIn} hold.

Until now, the questions (i) and (ii) have been completely answered
only for those particular mutual positions of the unperturbed
spectral sets $\sigma$ and $\Sigma$ where one of these sets lies in
a finite or infinite gap of the other one, say, $\sigma$ lies in a
gap of $\Sigma$. For such mutual positions,
\begin{equation}
\label{sin2t} c_*=\text{\small$\frac{1}{2}$}\quad\text{and}\quad
M(x)=\text{\small$\frac{1}{2}$}\arcsin(2x).
\end{equation}
This result is contained in the Davis-Kahan $\sin2\theta$ theorem
(see \cite{DK70}).

In the general case where no assumptions are done on the mutual
position of $\sigma$ and $\Sigma$, except for condition
\eqref{Vd12In}, the best available answers to the questions (i) and
(ii) are based on the bound
\begin{equation}
\label{s2t-g} \theta(\fP,\fQ)\leq
\text{\small$\frac{1}{2}$}\arcsin\text{\small$\frac{\pi\|V\|}{d}$}
\quad\text{if\,\,}\
\|V\|\leq\text{\small$\frac{1}{\pi}$}\,d
\end{equation}
proven in  \cite{AM-CAOT-2013} and called there the generic
$\sin2\theta$ estimate. The bound \eqref{s2t-g} remains the
strongest known bound for $\theta(\fP,\fQ)$ whenever
$\|V\|\leq\text{\small$\dfrac{4}{\pi^2+4}$}\,d$ (see \cite[Remark
4.4]{AM-CAOT-2013}; cf. \cite[Corollary 2]{Se2013-1}).

In \cite{AM-CAOT-2013}, it has been shown that the bound
\eqref{s2t-g} can also be used to obtain estimates of the form
\eqref{Mquest} for $\|V\|>\frac{1}{\pi}d$. To this end, one
introduces the operator path $H_t=A+tV$, $t\in[0,1]$, and chooses a
set of points
\begin{equation}
\label{points} 0=t_0<t_1<t_2<\ldots<t_n=1
\end{equation}
in such a way that
\begin{equation}
\label{cAM}
\frac{(t_{j+1}-t_j)\|V\|}{\dist\bigl(\omega_{t_j},\Omega_{t_j}\bigr)}\leq\frac{1}{\pi},
\end{equation}
where $\omega_t$ and $\Omega_t$ denote the disjoint spectral
components of $H_t$ originating from $\sigma$ and $\Sigma$,
respectively; $\omega_t=\spec(H_t)\cap\cO_{d/2}(\sigma)$ and
$\Omega_t=\spec(H_t)\cap\cO_{d/2}(\Sigma)$. Applying the
estimate \eqref{s2t-g} to the maximal angle between the spectral
subspaces $\Ran(\sE_{H_{t_j}}(\omega_{t_j}))$ and
$\Ran(\sE_{H_{t_{j+1}}}(\omega_{t_{j+1}}))$ of the corresponding
consecutive operators $H_{t_j}$ and $H_{t_{j+1}}$ and using, step by
step, the triangle inequality for the maximal angles (see
\cite{Brown}; cf. \cite[Lemma 2.15]{AM-CAOT-2013}) one arrives at the
optimization problem
\begin{equation}
\label{Opt}
\arcsin\bigl(\|P-Q\|\bigr)\leq\frac{1}{2}\,\,\,\inf_{n,\,\{t_i\}_{i=0}^n
}\,\,\sum_{j=0}^{n-1} \arcsin\frac{\pi
(t_{j+1}-t_j)\|V\|}{\dist\bigl(\omega_{t_j},\Omega_{t_j}\bigr)},
\end{equation}
over $n\in\mathbb{N}$ and $\{t_i\}_{i=0}^n$ chosen accordingly to
\eqref{points} and \eqref{cAM}. Taking into account that
$$
\dist\bigl(\omega_{t_j},\Omega_{t_j}\bigr)\geq d-2\|V\|t_j,
$$
from \eqref{Opt} one then deduces the bound
\begin{equation}
\label{Optg} \theta(\fP,\fQ)\leq M_{\rm
gen}\left(\text{\small$\frac{\|V\|}{d}$}\right)
\end{equation}
with the estimating function $M_{\rm gen}(x)$,
$x\in[0,\frac{1}{2})$, given by
\begin{equation}
\label{Mgen} M_{\rm
gen}(x)=\frac{1}{2}\,\,\,\inf_{n,\,\{\varkappa_i\}_{i=0}^n
}\,\,\sum_{j=0}^{n-1} \arcsin\frac{\pi
(\varkappa_{j+1}-\varkappa_j)}{1-2\varkappa_j},
\end{equation}
where the points
\begin{equation}
\label{pointx}
0=\varkappa_0<\varkappa_1<\varkappa_2<\ldots<\varkappa_n=x
\end{equation}
should be such that
$$
\frac{\varkappa_{j+1}-\varkappa_j}{1-2\varkappa_j}\leq\frac{1}{\pi}.
$$
An explicit expression for the function $M_{\rm gen}$ has been found
by Seelmann in \cite[The\-o\-rem 1]{Se2013-2}. From  \cite[Theorem
1]{Se2013-2} it also follows that the generic optimal constant $c_*$
in \eqref{Vcsd} satisfies inequalities
$$
c_{_{\rm S}}\leq c_*\leq\frac{1}{2},
$$
where
\begin{equation}
\label{cS} c_{_{\rm
S}}=\text{\small$\frac{1}{2}-\frac{1}{2}\biggl(1-\frac{\sqrt{3}}{\pi}\biggr)^3$}
=0.454839\ldots.
\end{equation}
The earlier results from \cite{AM-CAOT-2013}, \cite{KMMa}, and
\cite{MS2013-JRAM} concerning the generic bound \eqref{Mquest} might
be of interest, too.

The questions like (i) and (ii) have been addressed as well in the case
of off-diagonal perturbations. Recall that  a bounded operator $V$
is said to be off-diagonal with respect to the partition
$\spec(A)=\sigma\cup\Sigma$ of the spectrum of $A$ with
$\sigma\cap\Sigma=\emptyset$ if $V$ anticommutes with the difference
$P-P^\perp$ of the spectral projections $P=\sE_A(\sigma)$ and
$P^\perp=\sE_A(\Sigma)$, that is, if\,
$$
V(P-P^\perp)=-(P-P^\perp)V.
$$

When considering an off-diagonal Hermitian perturbation, one should
take into account that conditions ensuring the disjointness of the
respective perturbed spectral components $\omega$ and $\Omega$
originating from $\sigma$ and $\Sigma$ are much weaker than the
condition \eqref{Vd12In}. In particular, if the sets $\sigma$ and
$\Sigma$ are subordinated, say $\max(\sigma)<\min(\Sigma)$, then for
any (arbitrarily large) $\|V\|$ no spectrum of $H=A+V$ enters the
open interval between $\max(\sigma)$ and $\min(\Sigma)$ (see, e.g.,
\cite[Remark 2.5.19]{Tretter2008}). In such a case the maximal angle
$\theta(\fP,\fQ)$ between the unperturbed and perturbed spectral
subspaces $\fP$ and $\fQ$ admits a sharp bound of the form
\eqref{Mquest} with
\begin{equation}
\label{t2t} M(x)=\text{$\frac{1}{2}$}\arctan(2x),\quad
x\in[0,\infty).
\end{equation}
This is the consequence of the celebrated Davis-Kahan $\tan2\theta$
theorem \cite{DK70} (also, cf. the extensions of the $\tan2\theta$
theorem in \cite{GKMV2013,KMM5,MotSel}).

If it is known that the set $\sigma$ lies in a finite gap of the set
$\Sigma$ then the disjointness of the perturbed spectral components
$\omega$ and $\Omega$ is guaranteed by the (sharp) condition
$\|V\|<\sqrt{2}\,d$. The same condition is optimal for the bound
\eqref{tlp2} to hold. Both these results have been established in
\cite{KMM3}. An explicit expression for the best possible function
$M$ in the corresponding estimate \eqref{Mquest},
$$
M(x)=\arctan x,\quad x\in[0,\sqrt{2}),
$$
was found in \cite{AM-IEOT-2012,MotSel}.

As for the generic case --- with no restrictions on the mutual position
of the spectral components $\sigma$ and $\Sigma$,  the condition
\begin{equation}
\label{V32} \|V\|<\text{\small$\frac{\sqrt{3}}{2}$}d
\end{equation}
is known to be optimal in order to ensure that the gaps between
$\sigma$ and $\Sigma$ do not close  under an off-diagonal $V$.
Moreover, under this condition for the perturbed spectral sets
$\omega$ and $\Omega$ we have the following enclosures:
$$
\omega\subset \cO_{\epsilon_V}(\sigma)\text{\,\, and
\,\,}\Omega\subset \cO_{\epsilon_V}(\Sigma)
$$
with
\begin{equation}
\label{rV}
\epsilon_V=\|V\|\tan\left(\text{\small$\frac{1}{2}$}\arctan\text{\small$\frac{2\|V\|}{d}$}\right)
<\text{\small$\frac{d}{2}$}
\end{equation}
and, hence,
\begin{equation}
\label{doff} \dist(\omega,\Omega)\geq d-2\epsilon_V>0.
\end{equation}
The corresponding proofs were given initially in
\cite[Theorem~1]{KMM4} for bounded $A$ and then in \cite[Proposition
2.5.22]{Tretter2008} for unbounded $A$. From \eqref{V32} it follows
that the optimal constant $c_*$ in the condition \eqref{Vcsd}
ensuring the strict inequality \eqref{tlp2} in the generic
off-diagonal case necessarily satisfies the upper bound
\begin{equation}
\label{csoff}
c_*\leq\text{\small$\frac{\sqrt{3}}{2}$}\quad\bigl(=0.866025\ldots\bigr).
\end{equation}
Now we employ the approach of \cite{AM-CAOT-2013} in order
to get a lower bound for the above constant $c_*$. To this end, we
simply apply the optimization estimate \eqref{Opt} to the
off-di\-a\-go\-nal per\-tur\-bations. Due to \eqref{doff}, for
the disjoint spectral components $\omega_{t_j}$ and $\Omega_{t_j}$
of the operator $H_{t_j}=A+t_j V$ we have
\begin{align*}
\dist\bigl(\omega_{t_j},\Omega_{t_j}\bigr)\geq &\, d-
2t_j\|V\|\tan\left(\text{\small$\frac{1}{2}$}\arctan\text{\small$\frac{2t_j\|V\|}{d}$}\right)\\
&= 2d-\sqrt{d^2+4 t_j^2\|V\|^2}.
\end{align*}
The estimate \eqref{Opt} then yields
\begin{equation}
\label{Optff} \theta(\fP,\fQ)\leq M_{\rm
off}\left(\text{\small$\frac{\|V\|}{d}$}\right)
\end{equation}
with the function $M_{\rm off}(x)$,
$x\in[0,\frac{\sqrt{3}}{2})$, given by
\begin{equation}
\label{Moff} M_{\rm
off}(x)=\text{\small$\frac{1}{2}$}\,\,\,\inf_{n,\,\{\varkappa_i\}_{i=0}^n
}\,\,\sum_{j=0}^{n-1} \arcsin\text{\small$\frac{\pi
(\varkappa_{j+1}-\varkappa_j)}{2-\sqrt{1+4 \varkappa_j^2}}$},
\end{equation}
where $\varkappa_0=0$, $\varkappa_n=x$, and the remaining points
$\varkappa_j$, $j=1,2,\ldots,n-1$, should satisfy inequalities
$$
0<\text{\small$\frac{\varkappa_{j+1}-\varkappa_j}{2-\sqrt{1+4
\varkappa_j^2}}$}\leq\frac{1}{\pi}.
$$

We have only performed a partial numerical optimization of the
r.h.s. term in \eqref{Moff} restricting ourselves to the case where
the final function is smooth. As a result, our numerical
approximation $\widetilde{M}_{\rm off}$ for the estimating function
$M_{\rm off}$ for sure satisfies the bound
\begin{equation}
\label{MgM} \widetilde{M}_{\rm off}(x)\geq M_{\rm off}(x)
\quad\text{for all
\,}x\in\bigl[0,\text{\scriptsize$\frac{\sqrt{3}}{2}$}\bigr).
\end{equation}
The numerical function $\widetilde{M}_{\rm off}(x)$ is plotted in
Fig.\,\ref{fig1} along with the two previously known estimating
functions
$$
M_{\rm KMM}(x)=\arcsin\left(\min\left\{\,1,\,\,\text{\small$\frac{\pi\,
x}{3-\sqrt{1+4x^2}}$}\right\}\right),\quad
x\in\bigl[0,\text{\scriptsize$\frac{\sqrt{3}}{2}$}\bigr),
$$
from \cite[Theorem 2]{KMM4} and
$$
M_{\rm MS}(x)=\arcsin\left(\min\left\{\,1,\,\,\text{\normalsize$\frac{\pi}{2}\int_0^x
\frac{d\tau}{2-\sqrt{1+4\tau^2}}$}\right\}\right), \quad
x\in\bigl[0,\text{\scriptsize$\frac{\sqrt{3}}{2}$}\bigr),
$$
from \cite[Theorem 3.3]{MS2013-JRAM} that both serve as $M$ in the
bound \eqref{Mquest} for the case of off-di\-a\-go\-nal perturbations. For
convenience of the reader, in the plot we divide all the three
functions $M_{\rm KMM}$, $M_{\rm MS}$, and $\widetilde{M}_{\rm off}$
by ${\pi}/{2}$.

\begin{figure}
\centerline{\includegraphics[angle=0,width=0.92\textwidth]{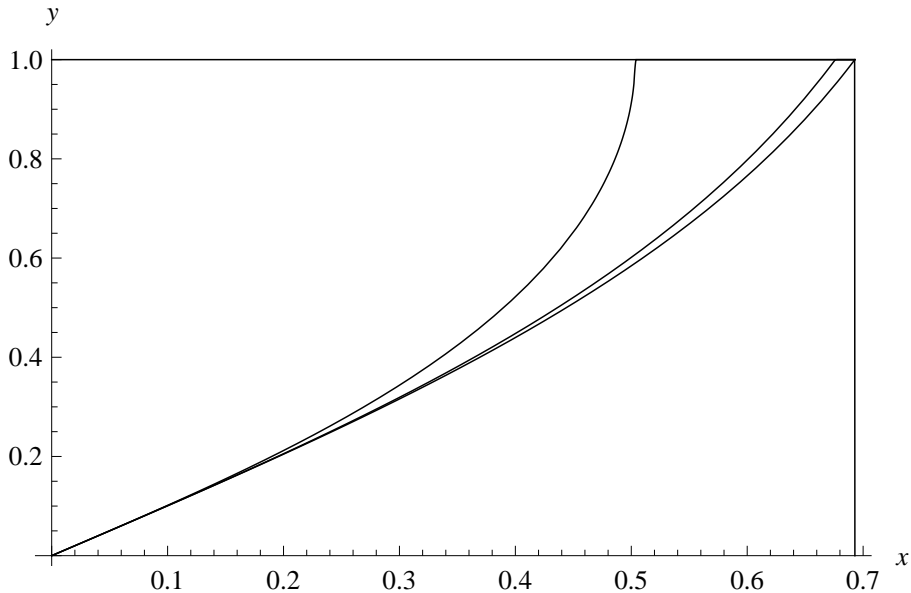}}
\caption{Graphs of the functions $\frac{2}{\pi}M_{\rm KMM}(x)$,
$\frac{2}{\pi}M_{\rm MS}(x)$, and the numerical approximation
$\frac{2}{\pi}\widetilde{M}_{\rm off}(x)$ for
$\frac{2}{\pi}M_{\rm off}(x)$ while its value does not exceed 1.
The upper curve depicts the graph of $\frac{2}{\pi}M_{\rm KMM}(x)$,
the intermediate curve is the graph of $\frac{2}{\pi}M_{\rm MS}(x)$,
and the lower curve represents the graph of
$\frac{2}{\pi}\widetilde{M}_{\rm off}(x)$.} \label{fig1}
\end{figure}

For the (unique) numerical solution $x=\widetilde{c}_{\rm off}$ of
the equation $\widetilde{M}_{\rm off}(x)={\pi}/{2}$ within the
interval $[0,\frac{\sqrt{3}}{2})$ we obtain
\begin{equation}
\label{cAMofft} \widetilde{c}_{\rm off}=0.692834\ldots.
\end{equation}
Since the function $\widetilde{M}_{\rm off}$ is monotonous and
inequality \eqref{MgM} holds, the number $\widetilde{c}_{\rm off}$
is an approximation to the exact solution $x=c_{\rm
off}>\widetilde{c}_{\rm off}$ of the equation $M_{\rm
off}(x)={\pi}/{2}$. Therefore, we arrive at the new lower bound
\begin{equation}
\label{cAMoff} c_*>0.692834
\end{equation}
for the optimal constant $c_*$ in the condition \eqref{Vcsd}
ensuring the subspace variation estimate \eqref{tlp2} in the generic
off-diagonal subspace perturbation problem. The bound \eqref{cAMoff}
is stronger than the corresponding best previously published bound
$c_*>0.67598$ from \cite{MS2013-JRAM}. Furthermore, we have
inequalities
\begin{equation}
\label{beb}
M_{\rm off}(x)\leq \widetilde{M}_{\rm off}(x)<  M_{\rm MS}(x) \quad
\text{for any \,} x\in(0,\widetilde{c}_{\rm off})
\end{equation}
which show that already the approximate estimating function
$\widetilde{M}_{\rm off}$ provides a bound of the form
\eqref{Mquest} that is stronger than the best known bound
(with the function $M_{\rm MS}$) from \cite{MS2013-JRAM}.

\section{Applications to few-body bound-state problems}
\label{Sec-FB}

From now on, we assume that the ``unperturbed'' Hamiltonian $A$ has
the form $A=H_0+V_0$ where $H_0$ is the kinetic energy operator of
an $N$-particle system in the c.m. frame and the potential $V_0$
includes only a part of the interactions that are present in the
system (say, only two-body forces). The perturbation $V$ describes
the remaining part of the interactions (say, three-body forces if
$N=3$; it may also describe the effect of external fields). We
consider the case where $V$ is a bounded operator. Of course, both
$A$ and $V$ are assumed to be Hermitian. In order to apply the
abstract results mentioned in the previous section, one only needs
to know the norm of the perturbation $V$ and a very basic stuff on
the spectrum of the operator~$A$.

Examples 3.1 and 3.2 below are borrowed from \cite{M-FBS-2014}.
\smallskip

The first of the examples represents a simple illustration of the
Davis-Kahan $\sin2\theta$ and $\tan2\theta$ theorems \cite{DK70}.%
\smallskip

\textit{Example 3.1} Suppose that {$E_0$} is the ground-state (g.s.)
energy of the Hamiltonian $A$. Also assume that the eigenvalue $E_0$
is simple and let $|\psi_0\rangle$ be the g.s.\,wave function, i.e.
$A|\psi_0\rangle=E_0|\psi_0\rangle$, $\|\psi_0\|=1$. Set
$\sigma=\{E_0\}$, $\Sigma=\spec(A)\setminus\{E_0\}$ and
$d=\dist(\sigma,\Sigma)=\min(\Sigma)-E_0$ (we notice that the set
$\Sigma$ is not empty since it should contain at least the essential
spectrum of $A$). If $V$ is such that the condition \eqref{Vd12In}
holds, then the g.s. energy $E'_0$ of the total Hamiltonian $H=A+V$
is again a simple eigenvalue, with a g.s. wave function
$|\psi'_0\rangle$, $\|\psi'_0\|=1$. The eigenvalue $E'_0$ lies in
the closed $\|V\|$-neighborhood of the g.s. energy $E_0$, i.e.
$|E_0-E'_0|\leq\|V\|$. The corresponding spectral projections
$P=\sE_A(\sigma)$ and $Q=\sE_H(\omega)$ of $A$ and $H$ associated
with the one-point spectral sets $\sigma=\{E_0\}$ and
$\omega=\{E'_0\}$ read as $P=|\psi_0\rangle\langle\psi_0|$ and
$Q=|\psi'_0\rangle\langle\psi'_0|$. One verifies by inspection
that\,
$$
\arcsin\bigl(\|P-Q\|\bigr)=\arccos|\langle\psi_0|\psi'_0\rangle|.
$$
Surely, this means that the maximal angle\, $\theta(\fP,\fQ)$
\,between the one-dimensional spectral subspaces
$\fP=\Ran(P)=\mathop{\rm span}(|\psi_0\rangle)$ and
$\fQ=\Ran(Q)=\mathop{\rm span}(|\psi'_0\rangle)$ is nothing but the
angle between the g.s. vectors $|\psi_0\rangle$ and
$|\psi'_0\rangle$. Then the Davis-Kahan $\sin2\theta$ theorem
implies (see \eqref{Mquest} and \eqref{sin2t}) that\,
$$
\arccos|\langle\psi_0|\psi'_0\rangle|\leq
\text{\small$\frac{1}{2}$}\arcsin\text{\small$\frac{2\|V\|}{d}$}.
$$
This bound on the rotation of the ground state means, in particular,
that, under the condition \eqref{Vd12In}, the angle between
$|\psi_0\rangle$ and $|\psi'_0\rangle$ can never exceed~$45^\circ$.

If, in addition, the perturbation $V$ is off-diagonal with respect
to the partition $\spec(A)=\sigma\cup\Sigma$ then for any
(arbitrarily large) {$\|V\|$} no spectrum of $H$ is present in the
gap between the g.s. energy $E_0$ and the remaining spectrum
$\Sigma$ of $A$. Moreover, there are the following sharp universal
bounds for the perturbed g.s. energy $E'_0$:
$$
\mbox{$E_0-\epsilon_V\leq E'_0\leq E_0$},
$$
(see \cite[Lemma 1.1]{KMM4} and \cite[Proposition
2.5.21]{Tretter2008}).  In this case, the Davis-Kahan $\tan2\theta$
theorem \cite{DK70} implies (see \eqref{Mquest} and \eqref{t2t})
that
$$
\arccos|\langle\psi_0|\psi'_0\rangle|\leq
\text{\small$\frac{1}{2}$}\arctan\text{\small$\frac{2\|V\|}{d}$}
<\text{\small$\frac{\pi}{4}$}.
$$

With a minimal change, the same consideration may be extended to the
case where the initial spectral set $\sigma$ consists of the $n+1$
lowest binding energies $E_0<E_1<\ldots< E_n$, $n\geq 1$,
of $A$. We only underline that if $V$ is off-diagonal than for any
$\|V\|$ the perturbed spectral set $\omega$ of $H=A+V$ originating
from $\sigma$ will necessarily be confined in the interval
$[E_0-\epsilon_V,E_{n}]$ where the shift $\epsilon_V$ is given by
\eqref{rV}; the interval $\bigl(E_n,\min(\Sigma)\bigr)$ will contain
no spectrum of $H$. Furthermore, the $\tan2\theta$-like estimates
for the maximal angle between the spectral subspaces
$\fP=\Ran\bigl(\sE_A(\sigma)\bigr)$ and
$\fQ=\Ran\bigl(\sE_H(\omega)\bigr)$ may be done even for some
unbounded $V$ (but, instead of $d$ and $\|V\|$, those estimates
involve quadratic forms of $A$ and $V$), see \cite{GKMV2013,MotSel}.
\smallskip

Along with the $\sin2\theta$ theorem, the next example illustrates
the $\tan\theta$ bound from \cite{AM-IEOT-2012,MotSel}.

\smallskip

\textit{Example 3.2} Suppose that
${\sigma=\{E_{n+1},E_{n+2},\ldots,E_{n+k}\}},\, {n\geq 0,\,\, k\geq
1},$ is a set formed by the consecutive binding energies of {$A$}
and $\Sigma=\spec(A)\setminus\sigma=\Sigma_-\cup\Sigma_+$ where
$\Sigma_-$ is the increasing sequence of the energy levels $E_0,
E_1,\ldots,E_n$ of $A$ that lie lower $\min(\sigma)$; $\Sigma_+$
denotes the remainder of the spectrum of $A$, that is,
$\Sigma_+=\spec(A)\setminus(\sigma\cup\Sigma_-)$. Under condition
\eqref{separIn}, this assumption means that the set $\sigma$ lies in
the finite gap $\bigl(\max(\Sigma_-),\min(\Sigma_+)\bigr)$ of the
set $\Sigma$. If one only assumes for $V$ the norm bound
\eqref{Vd12In} and makes no assumptions on the structure of $V$,
then not much can be said about the location of the perturbed
spectral sets $\omega$ and $\Omega$, except for \eqref{omOMd}.
However the Davis-Kahan $\sin2\theta$ theorem \cite{DK70} still well
applies and, thus, one has the bound
$$
\theta(\fP,\fQ)\leq\text{\small$\frac{1}{2}$}
\arcsin\text{\small$\frac{2\|V\|}{d}$}<\text{\small$\frac{\pi}{4}$}.
$$

Much stronger conclusions are done if $V$ is off-diagonal with
respect to the partition $\spec(A)=\sigma\cup\Sigma$.  In Section 2,
it was already mentioned that for off-diagonal $V$ the
gap-non-closing condition is of the form $\|V\|<\sqrt{2}d$ (and even
a weaker but somewhat more detailed condition $\|V\|<\sqrt{dD}$\,
with\, $D=\min(\Sigma_+)-\max(\Sigma_-)$\, is admitted, see
\cite{KMM3,MotSel}). In this case the lower bound for the spectrum
of $H=A+V$ reads as $E_0-\epsilon_V$ where the maximal possible
energy shift $\epsilon_V$, $\epsilon_V<d$, is given again by
\eqref{rV}. Furthermore, the perturbed spectral set $\omega$ is
confined in the interval $[E_{n+1}-\epsilon_V,E_{n+k}+\epsilon_V]$,
while the open intervals $(E_n,E_{n+1}-\epsilon_V)$ and
$\bigl(E_{n+k}+\epsilon_V,\min(\Sigma_+)\bigr)$ contain no spectrum
of $H$. For tighter enclosures for the perturbed spectral sets
$\omega$ and $\Omega$ involving the the gap length\, $D$,\, we refer
to \cite{KMM3,KMM4,Tretter2008}. In the case under consideration,
the sharp bound for the size of rotation of the spectral subspace
$\fP=\Ran\bigl(\sE_A(\sigma)\bigr)$ to the spectral subspaces
$\fQ=\Ran\bigl(\sE_H(\omega)\bigr)$ is given by the a priori
$\tan\theta$ theorem (see \cite[Theorem 1]{AM-IEOT-2012}; cf.
\cite[Theorem 2]{MotSel}):
$$
\theta(\fP,\fQ)\leq\arctan\text{\small$\frac{\|V\|}{d}$}<\arctan\sqrt{2}.
$$
If the gap length $D$\, is known and $\|V\|<\sqrt{dD}$, then a
stronger but more detailed estimate for $\theta(\fP,\fQ)$ is
available (see \cite[Theorem 4.1]{AM-IEOT-2012}).
\smallskip

\textit{Example 3.3} models the generic spectral disposition. Assume
that the binding energies of $A$ are numbered in the increasing
order, $E_0<E_1<\ldots<E_n<\ldots$, and
$\sigma=\{E_0,E_2,\ldots,E_{2k}\}$ is formed of the first $k+1$,
$k\geq 1$, binding energies with even numbers. Let
$\Sigma=\spec(A)\setminus\sigma$ and, thus, $\Sigma$ contains the
first $k$ binding energies $E_1,E_3,\ldots,E_{2k-1}$ with the odd
numbers, as well as the remaining point spectrum and the essential
spectrum of $A$. If $d=\dist(\sigma,\Sigma)>0$ and $\|V\|<c_{_{\rm
S}}d$ with $c_{_{\rm S}}$ given by \eqref{cS}, then for the maximal
angle $\theta(\fP,\fQ)$ between the corresponding unperturbed and
perturbed spectral subspaces $\fP=\Ran\bigl(\sE_A(\sigma)\bigr)$ and
$\fQ=\Ran\bigl(\sE_H(\omega)\bigr)$ we have the bound \eqref{Optg}.

If, in addition, the perturbation $V$ is off-diagonal with respect
to the partition $\spec(A)=\sigma\cup\Sigma$ then the disjointness
of the perturbed spectral components $\omega$ and $\Omega$ is
guaranteed by the weaker requirement $\|V\|<\frac{\sqrt{3}}{2}d$. In
this case $\omega\subset\cO_{\epsilon_V}(\sigma)$ and
$\Omega\subset\cO_{\epsilon_V}(\Sigma)$ where $\epsilon_V$ is given
by \eqref{rV}. Furthermore, if $\|V\|<\widetilde{c}_{\rm off}\,d$
where $\widetilde{c}_{\rm off}$ is the solution \eqref{cAMofft} of
the equation $\widetilde{M}_{\rm off}(x)={\pi}/{2}$, then one
can apply the bound \eqref{Moff}.
\smallskip

Examples 3.1\,--\,3.3 show how one may obtain a bound on
variation of the spectral subspace prior to any real calculations
for the total Hamiltonian $H$. In order to get such a bound, only
the knowledge of the values of $d$ and $\|V\|$ is needed.
Furthermore, if $V$ is off-diagonal, by using just these two
quantities one can also provide the stronger estimates (via
$\epsilon_V$) for the binding energy shifts.
\medskip

\noindent\textbf{Acknowledgments.} This work was supported by the
Deutsche Forschungsgemeinschaft, by the Heisenberg-Landau Program,
and by the Russian Foundation for Basic Research.

\end{document}